\documentclass[twocolumn,prb,superscriptaddress]{revtex4-2}

\usepackage{float}
\usepackage{graphicx}
\usepackage{amssymb,amsmath}
\usepackage{bm}% bold math
\usepackage{dcolumn}

\usepackage{hyperref}
\hypersetup{backref,colorlinks=true,breaklinks,urlcolor=blue,linkcolor=blue,citecolor=blue}
\usepackage{color}

\usepackage[capitalise]{cleveref}
\usepackage[notrig]{physics}   % principal value \pv  also includes 
\usepackage{braket}
\usepackage[normalem]{ulem}

\begin{document}

\title{Benchmarking quantum master equations beyond ultraweak coupling}
\author{Camilo Santiago Tello Breuer}
\email{c.tellobreuer@campus.tu-berlin.de}
\affiliation{Institut f\"ur Theoretische Physik, Technische Universit\"at Berlin, Hardenbergstrasse 36, 10623 Berlin, Germany}
\author{Tobias Becker}
\email{tobias.becker@tu-berlin.de}
\affiliation{Institut f\"ur Theoretische Physik, Technische Universit\"at Berlin, Hardenbergstrasse 36, 10623 Berlin, Germany}
\author{Andr\'{e} Eckardt}
\email{eckardt@tu-berlin.de}
\affiliation{Institut f\"ur Theoretische Physik, Technische Universit\"at Berlin, Hardenbergstrasse 36, 10623 Berlin, Germany}

\begin{abstract}
Recently, Nathan and Rudner derived a Gorini-Kossakowski-Sudarshan-Lindblad master equation from the Redfield equation. 
The claim is that the level of approximation is equal to that of the Redfield equation. 
Here we benchmark the Nathan-Rudner equation (NRE) against the exact solution of a damped harmonic oscillator and compare its performance to that of the time-dependent Redfield equation (RE). We find that which of the equations performs better depends on the regime considered.
It turns out that the short-time dynamics is generally much better captured by the RE, whereas the NRE delivers results comparable to those of the rotating-wave approximation.
For the steady state, in the high-temperature limit the RE again performs better and its solution approaches the exact result for ultrahigh temperatures. Nevertheless, here also the NR equation constitutes a good approximation. 
In the low-temperature limit, in turn, the NRE provides a better approximation than the RE. For too strong coupling, here the RE might even fail completely by predicting unphysical behaviour.
\end{abstract}

\maketitle
\section{Introduction} 
All quantum systems are coupled to their environment. 
Only rarely, the idealization of an isolated quantum system is found to be a good approximation and coherent evolution can be observed beyond very short times (e.g., for experiments with ultracold atoms \cite{blochManybodyPhysicsUltracold2008}). 
In thermal equilibrium and for ultraweak system-bath coupling (defined by the system-bath coupling being small compared to the energy level splitting in the system), 
the state of the system is described by statistical mechanics. 
This is very efficient and requires only a few thermodynamic variables, like temperature or chemical potential.
However, beyond ultraweak coupling and away from equilibrium, the state of the system does depend on the details of the environment. 
Since an explicit treatment of the total system-bath compound is often neither of interest nor feasible, the general strategy is to derive a master equation for the reduced density matrix of the system.
Ideally, the solution of the master equation should be an accurate description of the impact of the environment on the dynamics and steady state of the system. 
A common approach for deriving a quantum master equation is the Born-Markov approximation giving rise to the Redfield master equation \cite{AGRedfield65,EFranciscoMLRonald92,breuer,weissQuantumDissipativeSystems2012,AMCastilloDRReichman15,Alicki18,BenattiCruscinskiFloreanini2022}. 
This equation has been shown to describe the state of the system rather accurately \cite{postivity}. It correctly captures the coherences of the density matrix (the off-diagonal elements in energy representation) in second order system-bath coupling \cite{thingna,ThingnaPRE13,postivity,Alicki22,TimofeevTrushechkin2022}.
Only the populations of the energy eigenstates (the diagonal elements of the density matrix), require more sophisticated methods already in second order, see, e.g., Refs. ~\cite{JThingnaWJSheng2014,Groszkowski2022,beckerCanonicallyConsistentQuantum2022}. 

However, the Redfield equation has the disadvantage that it is not of Gorini-Kossakowski-Sudarshan-Lindblad (GKSL) form. 
This implies on the one hand that it might lead to unphysical violations of positivity (which are, however, significant only outside the range of the validity of the Born-Markov approximation) \cite{VRomero1989,ASuarezIOppenheim92,PPechukas94,EGevaERosenman2000,AMCastilloDRReichman15,postivity}. 
On the other hand, it does not allow the use of standard quantum-trajectory simulations, where the coherent evolution of an ensemble of pure states is interrupted by random quantum jumps. 
The conventional approach relies on the GKSL form and cannot directly be applied to the Redfield equation. Therefore, it seems to be desired to find master equations of GKSL form that provide an accurate description of the coupling to the environment (Note that recently also alternative quantum-trajectory approaches have been described that allow to unravel time-local non-Lindblad master equations such as the Redfield equation \cite{DonvilMGinanneschi2022,beckerQuantumTrajectoriesTimeLocal2023,settimoGeneralizedrateoperatorQuantumJumps2024}, without requiriing an effective extension of the state space like previously proposed methods. However, in the long-time limit, they might require a larger number of trajectories as the conventional approach, as discussed in Ref.~\cite{beckerOptimalFormTimelocal}.).
For ultraweak coupling, this is achieved using the rotating-wave approximation (RWA), which gives rise to the quantum optical master equation \cite{breuer,HCarmicheal1999,CWGardiner00}. However, the assumption of ultraweak coupling is often challenged. 
For instance, in quantum many-body systems energy levels can become exponentially small with the system size.

Recently numerous different GKSL master equations have been proposed to approximate the Redfield equation  \cite{farinaOpenquantumsystemDynamicsRecovering2019,davidovicCompletelyPositiveSimple2020,nathan-rudner,tobias,davidovicGeometricarithmeticMasterEquation2022,dabbruzzoTimedependentRegularizationRedfield2023,dabbruzzoRecoveringCompletePositivity2023}.
Of special interest is the Nathan-Rudner equation (NRE) in Refs.~\cite{nathan-rudner,nathanQuantifyingAccuracySteady2024},  which is similar to the geometric-arithmetic master equation in Ref.~\cite{davidovicCompletelyPositiveSimple2020}. 
In their work Nathan and Rudner claim that the approximation that lead to this equation are consistent with the Born-Markov approach, such that it is as accurate as the Redfield equation.
This is very promising in view of the efficient simulation of open quantum systems beyond ultraweak coupling. 
The NRE has been studied in numerous works \cite{NR-comment,nathanResponseCommentUniversal2020,leePerturbativeSteadyStates2022,davidovicGeometricarithmeticMasterEquation2022,tupkarySearchingLindbladiansObeying2023} also in relation to time-dependent master equations \cite{dabbruzzoTimedependentRegularizationRedfield2023,dabbruzzoRecoveringCompletePositivity2023}. 
In this paper, we benchmark the Nathan-Rudner equation by applying it to an exactly solvable model, given by a damped harmonic oscillator, and by comparing its performance to that of the Redfield and the quantum optical master equations. 
For this purpose, we consider both the transient evolution of the system as well as its steady state.

This paper is organised as follows: 
In \cref{master}, we recapitulate the quantum master equations considered in this paper, the Redfield, the quantum optical and the Nathan-Rudner equation. These will later be applied to the damped harmonic oscillator introduced in \cref{damped} together with its exact Hu-Paz-Zhang master equation. 
In \cref{benchmark}, we benchmark both the transient dynamics and the steady state of the approximate master equations against the results of the exact master equation. 

\section{Master equations for weak system-bath coupling}
\label{master}
The quantum master equations considered here are obtained by combining second-order perturbation theory in the system-bath coupling with the Born and Markov approximations. 
This gives rise to the time-dependent Redfield equation.
Applying further approximations, one also obtains the time-independent Redfield equation, the Nathan-Rudner equation and the quantum-optical master equation. 
In this section, we briefly review these equations. 

For the microscopic derivation, we start from the total Hamiltonian of the full system-bath compound $\hat{H}_{\text{tot}} = \hat{H}_S + \hat{H}_B + \hat{H}_{SB}$, where $\hat{H}_S$ is the Hamiltonian of the open system and $\hat{H}_B$ is the Hamiltonian of the bath. 
The interaction Hamiltonian $\hat{H}_{SB}$ is written in the canonical form $\hat{H}_\mathrm{SB} = \sqrt{\gamma} \hat{S} \otimes \hat{B} $, e.g., by doing a Schmidt decomposition, where $\hat{S}$ and $\hat{B}$ act solely on the system and bath, respectively. 
The generalization to multiple coupling terms is straightforward, but not considered here for simplicity.
The dimensionless quantity $\gamma$ captures the relative strength of the interaction compared to the energy scales of the system. 
Let $t_0$ denote the time at which system and bath are coupled. 
That is, for time $t_0$ we assume a factorized initial state $\hat{\rho}_\mathrm{tot}(t_0) = \hat{\rho}(t_0) \otimes \hat \rho_\mathrm{B}$, with $\hat{\rho}_\mathrm{B} = e^{-\beta \hat H_\mathrm{B}}/Z_\mathrm{B}$ being the thermal equilibrium state of the bath at inverse temperature $\beta$. 

\subsection{Redfield equation}
\label{redfield}
The Redfield equation is obtained in second order of the system-bath coupling from the Born-Markov approximation. 
For the reduced density matrix $\hat{\rho}(t) = \tr_B (\hat{\rho}_\mathrm{tot}(t))$ of the system it reads \cite{AGRedfield65,EFranciscoMLRonald92,breuer,weissQuantumDissipativeSystems2012,AMCastilloDRReichman15,Alicki18,BenattiCruscinskiFloreanini2022}
\begin{align}
	\frac{d}{dt}\hat{\rho}_I(t) = -\int_{t_0}^{t}d\tau C(t-\tau) [\hat{S}_{t}, \hat{S}_{\tau} \hat{\rho}_I(t)]+\text{H.c.}
	\label{eq:start}
\end{align}
in the interaction picture, which is indicated by the label $I$.
Here, $C(t-\tau) \equiv \gamma\langle\hat{B}_t \hat{B}_{\tau}\rangle/\hbar^2$ denotes the bath correlation function and $\hat{S}_t = \exp(-i\hat{H}_St/\hbar)\hat{S}\exp(i\hat{H}_St/\hbar)$ and $\hat{B}_t = \exp(-i\hat{H}_Bt/\hbar)\hat{B}\exp(i\hat{H}_Bt/\hbar)$ are interaction picture operators. 
In the Schrödinger picture it takes the form
\begin{align}
	\begin{split}
		\frac{d}{dt}\hat{\rho}(t) =& -\frac{i}{\hbar}[\hat{H}_S, \hat{\rho}(t)]+\hat{S}\hat{\rho}(t) \hat{\mathbb{S}}^{\dagger}_{t} + \hat{\mathbb{S}}_{t}\hat{\rho}(t) \hat{S} \\&- \hat{S}\hat{\mathbb{S}}_{t}\hat{\rho}(t) - \hat{\rho}(t)\hat{\mathbb{S}}^{\dagger}_{t}\hat{S},
		\label{eq:Redfield}
	\end{split}
\end{align}
with $\hat{\mathbb{S}}_{t} \equiv \int_{t_0}^{t}d\tau C(\tau)\hat{S}_{-\tau}$. 
The last four terms give rise to both a Lamb-shift Hamiltonian and a dissipative contribution \cite{tobias,colla}. The Redfield equation is explicitly time-dependent. However, often the time-independent Redfield equation is considered by replacing $\hat{\mathbb{S}}_t \rightarrow \hat{\mathbb S}_\infty \equiv \int_{t_0}^\infty d\tau C(\tau)\hat{S}_{-\tau}$. This is justified for times $t-t_0$ that are large compared to the typical relaxation time of the bath correlation.

\subsection{Quantum optical master equation}
\label{sec:RWA}
For ultraweak system-bath coupling, we can perform a rotating-wave approximation (RWA), where we neglect off-diagonal coupling terms in the energy eigenbasis of the system, to obtain the quantum optical master equation \cite{breuer,weissQuantumDissipativeSystems2012,CWGardiner00}
\begin{align}
	\begin{split}
		&\frac{d}{dt}\hat{\rho}(t) = -\frac{i}{\hbar}[\hat{H}_S + \hat{\Lambda}^{\text{RWA}}, \hat{\rho}(t)] \\ &+ \sum_{lk}2G_t^r(\Delta_{lk})|S_{lk}|^2\left( \hat{L}_{lk}\hat{\rho}(t)\hat{L}_{lk}^{\dagger} -\frac{1}{2}\{\hat{L}_{lk}^{\dagger}\hat{L}_{lk}, \hat{\rho}(t)\} \right).
	\end{split}
\end{align}
Here, $G_t^r(\Delta) + i G_t^i(\Delta) \equiv \int_{0}^t d\tau C(\tau) e^{-i\Delta \tau}$ denotes the time-dependent coupling density, with real and imaginary parts $G_t^r(\Delta),\, G_t^i(\Delta)$, respectively. It also defines the tensor elements of the Redfield superoperator in the eigenbasis of the system. 
Often, especially for the dynamics approaching the steady state, we take the asymptotic generator with $G_\infty(\Delta)$, which is discussed in \cref{generator}. 
This will also be done in the following.
The Lamb-shift Hamiltonian is then diagonal in the eigenbasis of the system Hamiltonian $\hat{H}_S$,
\begin{align}
	\hat{\Lambda}^{\text{RWA}} =  \hbar\sum_{lk} G_\infty^i(\Delta_{lk})|S_{lk}|^2 \hat{L}_{lk}^{\dagger} \hat{L}_{lk} .
\end{align}
It, therefore, only shifts the eigenenergies of the system, but leaves the eigenstates unchanged.  

For thermal environments the steady-state solution of the quantum optical master equation is the canonical Gibbs state $\hat{\rho}_0 = \exp(-\beta \hat{H}_S)/Z_S$, with $Z_S = \tr_S(\exp(-\beta \hat{H}_S))$. Since this is independent of the coupling strength, it is only valid in the ultraweak coupling regime, where the coupling strength approaches zero. \cite{haenggi,thingna}.

\subsection{Nathan-Rudner equation}
\label{NR}
In this work, we focus on the master equation proposed by Nathan and Rudner in Ref.~\cite{nathan-rudner}. 
Like the quantum optical master equation, it is in GKSL form, but it is obtained from the Redfield equation by an approximation that is different from the RWA.
The claim is that, differently from the RWA, the error induced by this approximation is on the same order of magnitude as the error of the Redfield equation. 
The main idea is to introduce yet a second time integral on the right-hand side of \cref{eq:start} to identify an equation in GKSL form with one single jump operator and a new Lamb-shift Hamiltonian. In order to be able to do that, the authors explicitly consider $t_0=-\infty$, meaning that $t-t_0$ is large compared to the typical relaxation time of the bath correlation. 
In \cref{schroedinger}, we show that this approximation is equivalent to using the time-independent Redfield equation. 

To get the Nathan-Rudner equation, firstly, the bath correlation function is decomposed into a convolution integral
\begin{align}
	C(t-\tau) = \int_{-\infty}^{\infty}ds g(t-s)g(s-\tau), \label{eq:convolution}
\end{align}
which defines a new function $g(t)$. In practice, we will use the Fourier transform $h(\Delta)$ of this function, which, thanks to the convolution theorem, can be obtained by taking the square root of the Fourier transform of the bath correlation function $C(t)$ \cite{nathan-rudner}. Now, the Redfield \cref{eq:start} reads,
\begin{align}
	\frac{d}{dt}\hat{\rho}_I(t) = \int_{-\infty}^{\infty}d\tau \int_{-\infty}^{\infty}ds \mathcal{F}(t,s,\tau)[\hat{\rho}_I(t)],
\end{align}
with superoperator
\begin{equation}
 \mathcal{F}(t,s,\tau)[\hat{\sigma}] = \theta (t-\tau) g(t-s) \allowbreak g(s-\tau)[\hat{S}_{t}, \hat{S}_{\tau} \hat{\sigma}]+\text{H.c.}, 
 \end{equation}
 and $\theta (t)$ being the Heaviside step function. Integrating once from $t_1$ to $t_2$ results in
\begin{align}
	\hat{\rho}_I(t_2) - \hat{\rho}_I(t_1) = \int_{t_1}^{t_2}dt \int_{-\infty}^{\infty}d\tau \int_{-\infty}^{\infty}ds \mathcal{F}(t,s,\tau)[\hat{\rho}_I(t)].
\end{align}
Next, the integral is approximated according to
\begin{align}
\mathcal{F}(t,s,\tau) \approx \mathcal{F}(s,t,\tau).
\end{align}
By exchanging the variables $t$ and $s$, one obtains
\begin{align}
	\hat{\rho}_I(t_2) - \hat{\rho}_I(t_1) = \int_{t_1}^{t_2}ds \int_{-\infty}^{\infty}dt \int_{-\infty}^{\infty}d\tau \mathcal{F}(t,s,\tau)[\hat{\rho}_I(s)].
\end{align}
This step is described in detail in Ref.~\cite{nathan-rudner}, where the validity of the approximation is also discussed. The Nathan-Rudner equation is then obtained by performing a derivative with respect to $t_2$. We replace $\tau, \, t, \, t_2$ with $s', \, s, \, t$ respectively.
With the jump operator
\begin{align}
	\hat{L}_I = \int_{-\infty}^{\infty} ds g(t-s)\hat{S}_s , 
 \label{eq:L-I}
\end{align}
and the Lamb-shift Hamiltonian
\begin{align}
	\hat{\Lambda}_I^{N} = \frac{\hbar}{2i}\int_{-\infty}^{\infty} ds \int_{-\infty}^{\infty} ds'\hat{S}_s g(s-t) g(t-s')\hat{S}_{s'} \text{sgn}(s-s'),
\end{align}
it is possible to bring the equation into GKSL form,
\begin{align}
	\frac{d}{dt}\hat{\rho_I}(t) = -\frac{i}{\hbar}[\hat{\Lambda}_I^{N}, \hat{\rho}_I(t)] -\frac{1}{2}\{\hat{L}_I^{\dagger}\hat{L}_I, \hat{\rho}_I(t)\} + \hat{L}_I\hat{\rho}_I(t)\hat{L}_I^{\dagger}.
\end{align}
This is a remarkable result. Here $\theta (t) = \frac{1}{2} + \frac{1}{2} \text{sgn}(t)$ was used to separate the double integral into separable and inseparable parts, which represent the jump operator $\hat{L}_I$ and the Lambshift Hamiltonian $\hat{\Lambda}_I^{N}$, respectively. In the Schrödinger picture, the equation is given by
\begin{align}
	\frac{d}{dt}\hat{\rho}(t) = -\frac{i}{\hbar}[\hat{H}_S + \hat{\Lambda}^{N}, \hat{\rho}(t)] -\frac{1}{2}\{\hat{L}^{\dagger}\hat{L}, \hat{\rho}(t)\} + \hat{L}\hat{\rho}(t)\hat{L}^{\dagger}.
	\label{eq:NR}
\end{align}

For time-independent Hamiltonians it is convenient to use the representation of the Nathan-Rudner equation in the eigenbasis of the Hamiltonian of the system $\hat{H}_S$. The matrix elements of the operators $\hat{L}$ and $\hat{\Lambda}^{N}$ are then given by 
\begin{align}
L_{lk} = 2\pi h(\Delta_{lk})x_{lk}, \quad \Lambda_{lk}^{N} = \sum_{n}f(\Delta_{nl},\Delta_{kn})x_{ln}x_{nk}, \label{eq:NR-def}
\end{align}
where $\Delta_{lk} = \epsilon_l-\epsilon_k$ is the energy difference between the eigenstates $\ket{l}$ and $\ket{k}$. For thermal baths $h(\Delta)$ and $f(\Delta_1,\Delta_2)$ are defined as
\begin{align}
h(\Delta) &= \sqrt{\frac{1}{2\pi}\frac{J(\Delta)/\hbar}{e^{\beta \Delta}-1}}, 
\\ f(\Delta_1,\Delta_2) &= 2\pi \hbar \mathcal{P}\int_{-\infty}^{\infty}\frac{d\omega}{\omega} h(\omega + \Delta_1)h(\omega - \Delta_2),
\label{eq:f}
\end{align}
where $J(\Delta)$ is the spectral density (see \cref{generator}) and $\mathcal{P}$ denotes the principle value of the integral. 
Using convolution theorems, it can be shown that $h(\Delta)$ is the Fourier transform of $g(t)$ from \cref{eq:convolution}.

In Refs.~\cite{nathanTopologicalPhenomenaPeriodically,davidovicCompletelyPositiveSimple2020,davidovicGeometricarithmeticMasterEquation2022}, a similar master equation has been found by replacing the arithmetic mean in the Kossakowski matrix of the Redfield equation with a geometric mean. 
This approach leads to the same jump operator $\hat L_I$ in \cref{eq:L-I}.
The only difference between the geometric-arithmetic master equation and the NRE lies in the Lamb-shift Hamiltonian.
An alternative approach derives a Lindblad-form master equation from the dynamic structure factor \cite{vacchiniTestParticleQuantum2001}.

\section{Damped harmonic oscillator}
\label{damped}
In Ref.~\cite{nathan-rudner}, the authors described the regime of validity of their equation by introducing general error measures. In this work, we benchmark the master equation explicitly against the exact the solution of a damped harmonic oscillator. 
Furthermore, we compare with the Redfield and quantum optical master equation.

To this end we will use $H_S = \hbar \omega (\hat{a}^\dagger \hat{a} + 1/2) + G_{R} \hat{S}^2 $ as the Hamiltonian of the system, where $G_{R} = \int_0^\infty d\omega J(\omega)/\omega $ is the reorganization energy, which renormalizes the shifted energies in an open quantum system \cite{caldeira}.
The system is coupled to a thermal bath of harmonic oscillators $\hat{H}_B$ with coupling Hamiltonian, $\hat{H}_{SB} = \hat{x} \otimes \hat{B}$, where $\hat{x}$ is the displacement operator of the central oscillator.
\subsection{Nathan-Rudner equation} 
For an Ohmic spectral density with Drude cutoff, we are not able to find an analytical expression for the integral $f$ in \cref{eq:f}.
Thus, for a generic system, $f$ has to be calculated either for all relevant level splittings, which is done below, or it has to be approximated by interpolating it between values computed numerically on a sufficiently dense grid of arguments.
Compared to the computation of the Redfield generator, for which the coupling density can be expressed by analytic functions, the computation of the generator for the NRE, thus, requires a slightly larger numerical effort. 
Alternatively, one might also use the master equation proposed in Ref. \cite{davidovicGeometricarithmeticMasterEquation2022}, which is equivalent to the NRE, except for the fact that the Lamb-shift Hamiltonian takes a simpler form, which in fact directly corresponds to the Lamb-shift of the Redfield equation. 
For the harmonic oscillator to be considered here the energy levels are equidistant and the coupling operator $\hat x$ only couples states of energy difference $|\epsilon_l - \epsilon_k| = \hbar \omega$. 
Therefore, only three different integrals have to be calculated numerically, i.e, $f(-\hbar \omega, \hbar \omega)$, $f(\hbar \omega, -\hbar \omega) $ and $f(\hbar \omega, \hbar \omega)$. 

\subsection{Exact equation}
\label{exact}

\begin{figure}[t]
	\centering
	\includegraphics{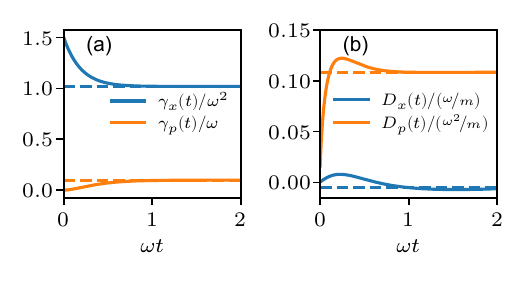}
	\caption{Time dependent coefficients of the exact Hu-Paz-Zhang master equation for $\gamma/\omega=0.1$, $E_c=5\hbar \omega$, $\beta\omega=1$.}
	\label{fig:time_dependent_coeff}
\end{figure}

To see how the NRE compares to the Redfield equation, the exact Hu-Paz-Zhang master equation for the harmonic oscillator with mass $m$ and frequency $\omega$ coupled to an Ohmic Drude bath with spectral density
\begin{align}
	J(\Delta) = \frac{\gamma \Delta/\pi}{1+(\Delta/E_c)^2} \label{eq:drude}
\end{align}
is considered, where $E_c$ is the cutoff energy.
For a factorized initial state with a bath in thermal equilibrium, the exact equation can be obtained ~\cite{karrlein,haake,hu},
\begin{align}
\begin{split}
\frac{d}{dt}\hat{\rho}(t) &= -\frac{i}{\hbar} \left[\frac{\hat{p}^2}{2m}+ \frac{m}{2}\gamma_x(t)\hat{x}^2, \hat{\rho}(t)\right]-i\frac{ \gamma_p(t)}{2\hbar}[\hat{x}, \{\hat{p}, \hat{\rho}(t)\}] \\ &+ \frac{m^2}{\hbar^2}D_p(t)[\hat{x},[\hat{p}, \hat{\rho}(t)]] - \frac{m^2}{\hbar^2}D_x(t)[\hat{x},[\hat{x}, \hat{\rho}(t)]]. \label{eq:Exact}
\end{split}
\end{align}
Here $\hat{p}$ is the momentum operator. 
The derivation of the four time-dependent coefficients $\gamma_x(t)$, $\gamma_p(t)$, $D_x(t)$ and $D_p(t)$ is outlined, e.g., in Refs.~\cite{GrabertWeiss84,HaakeReibold85,grabertQuantumBrownianMotion1988,HuPazZhang1992,Paz94,zerbeBrownianParametricQuantum1995,karrlein,JPiilo2004,HomaBernad2020}. 
The renormalized squared frequency $\gamma_x(t)$ and the relaxation strength $\gamma_p(t)$ are temperature independent. 
The temperature-dependent diffusion coefficients $D_x(t)$ and $D_p(t)$ result from quantum fluctuations.  
In the asymptotic regime, for $t\to \infty$, these coefficients approach constant values. 
This is analogous to the static limit discussed for the Redfield equation in \cref{generator}. 
For ultraweak coupling and large temperature, the time dependent coefficients are depicted in \cref{fig:time_dependent_coeff}. 
In the static limit, one obtains $\gamma_x\simeq \omega^2$, $\gamma_p \simeq \gamma$, $D_x\simeq 0$ and $D_p\simeq \gamma/m\beta$ (cf.~dashed lines in \cref{fig:time_dependent_coeff}).

For $\gamma$ approaching 0, the exact steady-state solution reduces to the canonical Gibbs state of the system \cite{thingna}. This means, that the steady-state solution of the Redfield equation is equal to the exact steady-state in the ultraweak coupling limit.
\section{Benchmarking against exact solution}
\label{benchmark}
In this section, we study the validity of the NRE by benchmarking it against the exactly solvable model of the damped harmonic oscillator and comparing the results to the Redfield equation and the quantum optical (or RWA) master equation.
All dynamical calculations are done for an initial pure state $\hat{\rho}_0 = \ketbra{\psi_0}{\psi_0}$ that is the equal superposition of the ground and the first excited state of the harmonic oscillator, $|\psi_0\rangle = (|0\rangle + |1\rangle)/ \sqrt{2}$ with $|1\rangle = \hat{a}^\dagger|0\rangle$, with $\hat{a} = \sqrt{m\omega/2\hbar} (\hat{x}+ i \hat{p}/m\omega)$. 
For our analysis, we use the full time-dependent exact and Redfield master equations. 
For a comparison to the approximative asymptotic time-independent generators, we refer to \cref{generator}.
The infinite-dimensional Hilbert space of the harmonic oscillator is truncated to the first 30 lowest energy eigenstates.
Temperatures and cutoff energies are given in units of $\hbar \omega$. 
Superscripts $X$ on variables and operators are used to refer to the different models, $X=E$ for the exact equation, $X=R$ for Redfield, and $X=N$ for Nathan-Rudner. 

\begin{figure}[b!]
	\hspace*{-0.6cm}
	\includegraphics[width=0.53\textwidth]{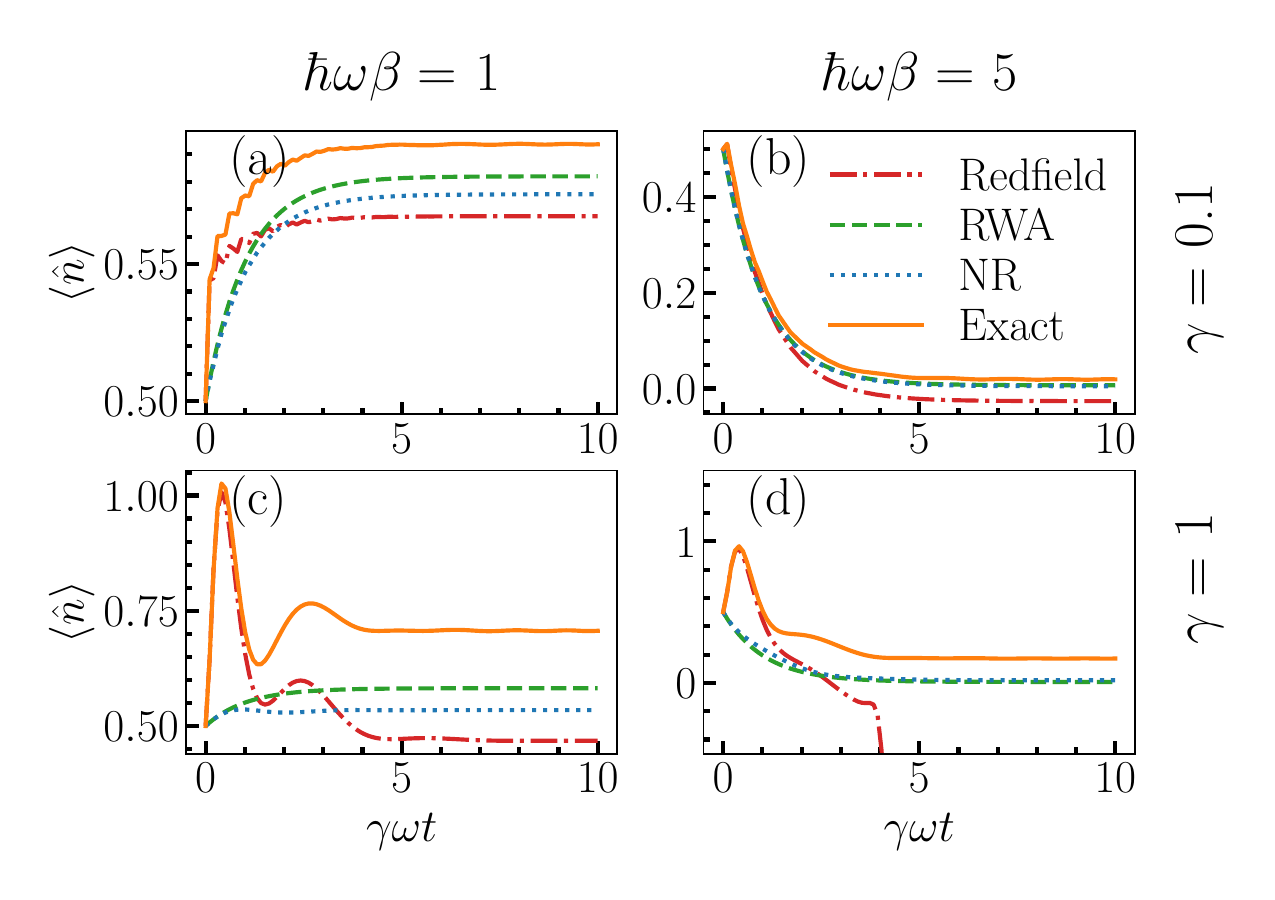}\\ %{Dynamics-hamiltonian-cutoff5}\\
	\caption[Examples for the dynamics of the expectation value of the Hamiltonian of the system using different values for $\gamma$ and $\beta$. All examples where calculated with $E_c = 5 \hbar \omega$. Values used for (a): $\gamma = 0.1$ and $\beta = 1$, (b): $\gamma = 0.1$ and $\beta = 5$, (c) $\gamma = 1$ and $\beta = 1$, (d) $\gamma = 1$ and $\beta = 5$.]{Examples for the dynamics of the expectation value of the number operator of the system using different values for $\gamma$ and $\beta$. All examples where calculated with $E_c = 5 \hbar \omega$.}
	\label{fig:hamiltonian-ex}
\end{figure}

\begin{figure}[t!]
	\hspace*{-0.6cm}
	\includegraphics[width=0.53\textwidth]{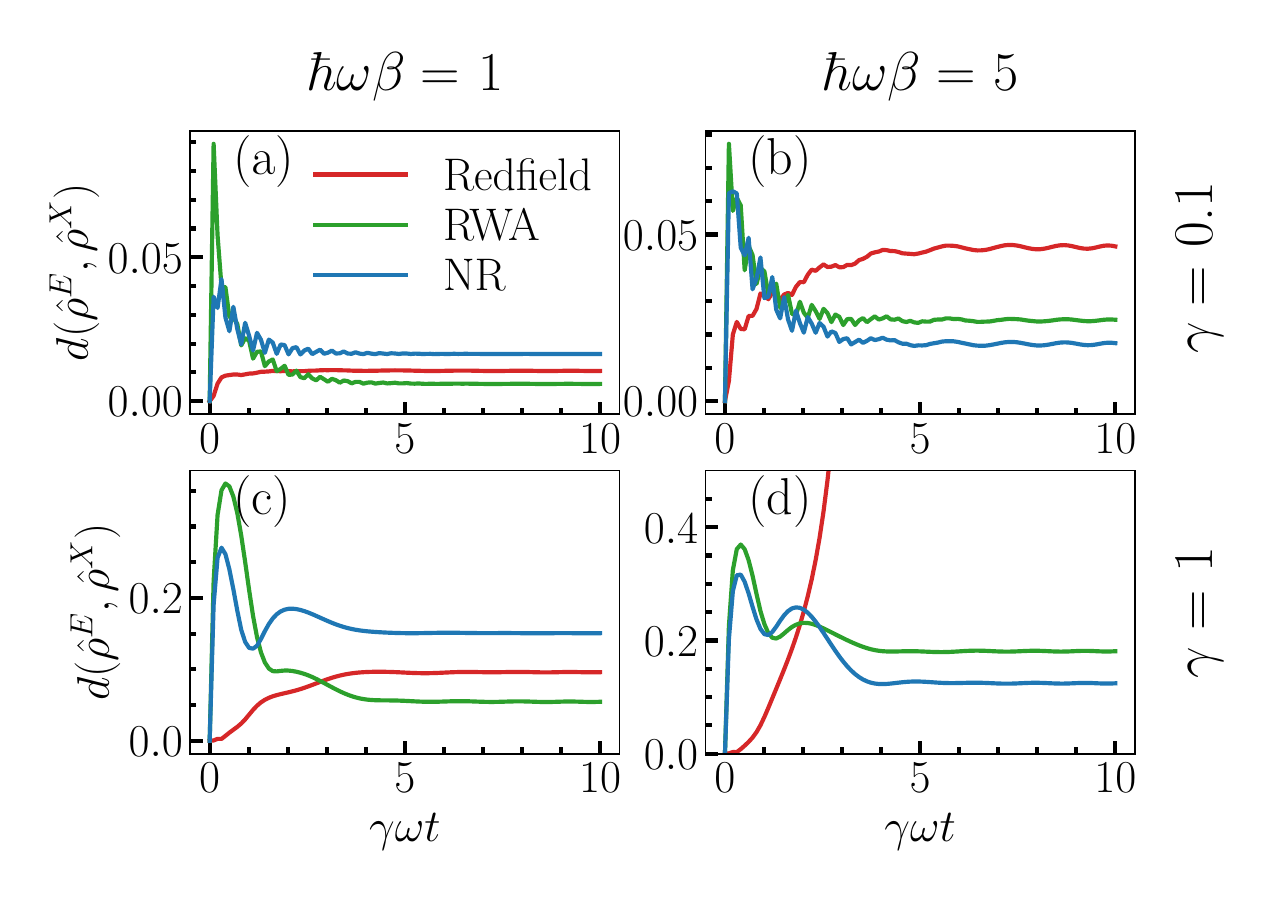}\\ %{Dynamics-tracedistance-cutoff5}\\
	\caption[Examples for the dynamics of the trace distance $d(\hat{\rho}^E, \hat{\rho}^X)$ between the approximate solutions and the exact solution using different values for $\gamma$ and $\beta$. All examples where calculated with $E_c = 5 \hbar \omega$. Values used for (a): $\gamma = 0.1$ and $\beta = 1$, (b): $\gamma = 0.1$ and $\beta = 5$, (c) $\gamma = 1$ and $\beta = 1$, (d) $\gamma = 1$ and $\beta = 5$.]{Examples for the dynamics of the trace distance $d(\hat{\rho}^E, \hat{\rho}^X)$ between the approximate solutions and the exact solution using different values for $\gamma$ and $\beta$. All examples where calculated with $E_c = 5 \hbar \omega$.}
	\label{fig:tracedistance-ex}
\end{figure}

\cref{fig:hamiltonian-ex,fig:tracedistance-ex}, give a first impression of the dynamics under the different master equations. 
They show the dynamics of the expectation value of the system's number operator $\langle \hat{n} \rangle = \langle \hat{a}^\dagger \hat{a} \rangle$ in \cref{fig:hamiltonian-ex}, and the trace distance $d(\rho^E, \rho^X)$ between the approximate solutions and the exact solution in \cref{fig:tracedistance-ex}, each for different inverse temperatures and coupling strengths.
The dynamics of $\langle \hat{n}\rangle$ in \cref{fig:hamiltonian-ex} indicate, that the solution of the Redfield equation follows the solution of the exact equation for a finite time on the order of $\gamma\omega t\sim 1$, while the Nathan-Rudner and RWA solutions deviate instantly from the exact dynamics. 
Looking at \cref{fig:tracedistance-ex}, this translates to a smaller trace distance between the Redfield solution and the exact solution as compared to the other master equations.
This observation is explained, most likely, by the fact that the NRE possesses a time-independent generator corresponding to the asymptotic long-time limit. 
Also the time-independent Redfield equation performs worse than the time-dependent one  (see \cref{fig:timeindependent} in \cref{generator}). 
However, unlike for the Redfield equation, no time-dependent equivalent of the NRE is known.

We also see in the Figures that at about $\gamma\omega t\sim 5$, the system approaches a steady state. Here the time-dependence of the generator becomes irrelevant. From the errors in \cref{fig:tracedistance-ex}, we infer that, while the Redfield equation performs slightly better for the larger temperature ($\hbar\omega\beta=1$), the NRE is superior for the lower temperature ($\hbar\omega\beta=5$). The Redfield solution even becomes unphysical for low temperature and strong system-bath coupling, giving rise to an unphysical negative average excitation number $\langle \hat{n} \rangle < 0$ and a trace distance larger than one. This is a well-known problem of the Redfield equation \cite{VRomero1989,ASuarezIOppenheim92,PPechukas94,EGevaERosenman2000,AMCastilloDRReichman15,postivity,deVaga}.
It does not occur for the NRE thanks to its GKSL form.

To get a better understanding of the differences between the master equations, in the following, we systematically analyse the relaxation dynamics of the different models. 
This is followed by an analysis of the steady state.

\subsection{Dynamics}
To systematically compare the dynamics of the different master equations, we average the trace distance $d(\rho^E, \rho^X)$  over the time interval $\gamma \omega t \in [0, \tau_R]$,
\begin{align}
	\overline{d(\rho^E, \rho^X)} = \frac{1}{\tau_R} \int_0^{\tau_R} d(\rho^E(t), \rho^X(t))\, dt.
	\label{eq:aver_trace_dist}
\end{align}
Here we have chosen the time interval with $\tau_R = 2/\gamma \omega$ based on the dynamics depicted in \cref{fig:hamiltonian-ex,fig:tracedistance-ex}. It covers the transient dynamics and excludes the steady state regime. Thus, it is roughly determined by the time scale on which the system relaxes to its steady state. 
The time-averaged distance $\overline{d(\rho^E, \rho^X)} $ defines the dynamical error and quantifies deviations from the exact solution during  transient evolution.
\begin{figure}[h]
	\hspace*{-0.6cm}
	\includegraphics[width=0.55\textwidth]{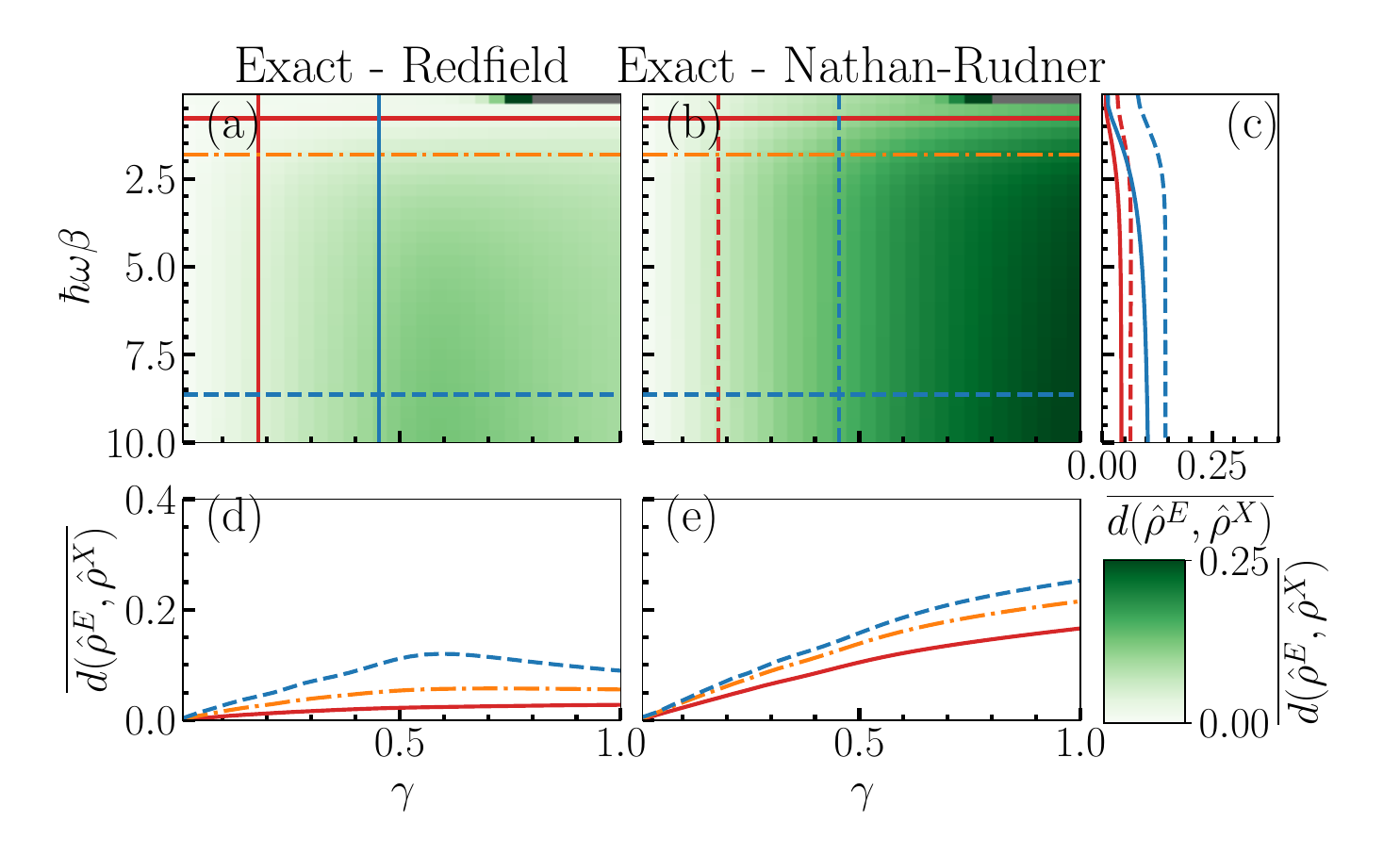}\\%{parameterscan-dynamic-redfield-full-cutoff5}\\
	\caption[Heatmaps of the average trace distances $d(\hat{\rho}^E,\hat{\rho}^X)$ during time evolution for cutoff $E_c = 5 \hbar \omega$ and varying values of coupling strength $\gamma$ and inverse temperature $\beta$]{Time average trace distance in the transient regime \cref{eq:aver_trace_dist} for cutoff $E_c = 5 \hbar \omega$. (a): 
	Heatmap of $\overline{d(\hat{\rho}^E,\hat{\rho}^R)}$ for a range of values of coupling strength $\gamma$ and inverse temperature $\beta$. The gray area indicates values of $\overline{d(\hat{\rho}^E,\hat{\rho}^X)} > 1$. (b): Equivalent Heatmap for $\overline{d(\hat{\rho}^E,\hat{\rho}^N)}$. (c) Plots of $\overline{d(\hat{\rho}^E,\hat{\rho}^X)}$ for specific values of $\gamma$. The corresponding $\gamma$ values for the graphs shown are marked in the heatmaps on the left by vertical lines. (d) and (e): Plots of $\overline{d(\hat{\rho}^E,\hat{\rho}^X)}$ for specific values of $\beta$. The corresponding $\beta$ values for the graphs shown are marked in the heatmaps above by horizontal lines.}
	\label{fig:maxtracedist-timev1}
\end{figure}
\cref{fig:maxtracedist-timev1} compares the dynamical errors of the Nathan-Rudner and Redfield equations as a function of inverse temperature $\beta$ and coupling strength $\gamma$.
For short time dynamics, the Redfield equation outperforms the NRE in the parameter regime shown in \cref{fig:maxtracedist-timev1} (a) and (b). 
The only exception being low temperatures and strong coupling strengths, where the performance is on the same order of magnitude (though for an even shorter time interval the Redfield equation would improve). 
This can be seen by the dotted blue lines in \cref{fig:maxtracedist-timev1} (d) and (e), where for fixed temperature the dynamical error is plotted as a function of the coupling strength $\gamma$. 
In \cref{fig:maxtracedist-timev1} (c), we can, in turn, observe that the errors vanish in the high-temperature (small $\hbar\omega\beta$) limit for the Redfield results, while this does not seem to be the case for the NRE (see also discussion below).

As discussed above, one of the reasons for the superior performance of the Redfield equation in this case seems to come from the fact that we use the time-dependent Redfield equation, whereas for the NRE, the starting point had to be the asymptotic (time-independent) Redfield equation. 
This is highlighted in \cref{generator}, where we compare the time-independet Redfield equation to the NRE.

\begin{figure}[h]
	\hspace*{-0.6cm}
	\includegraphics[width=0.55\textwidth]{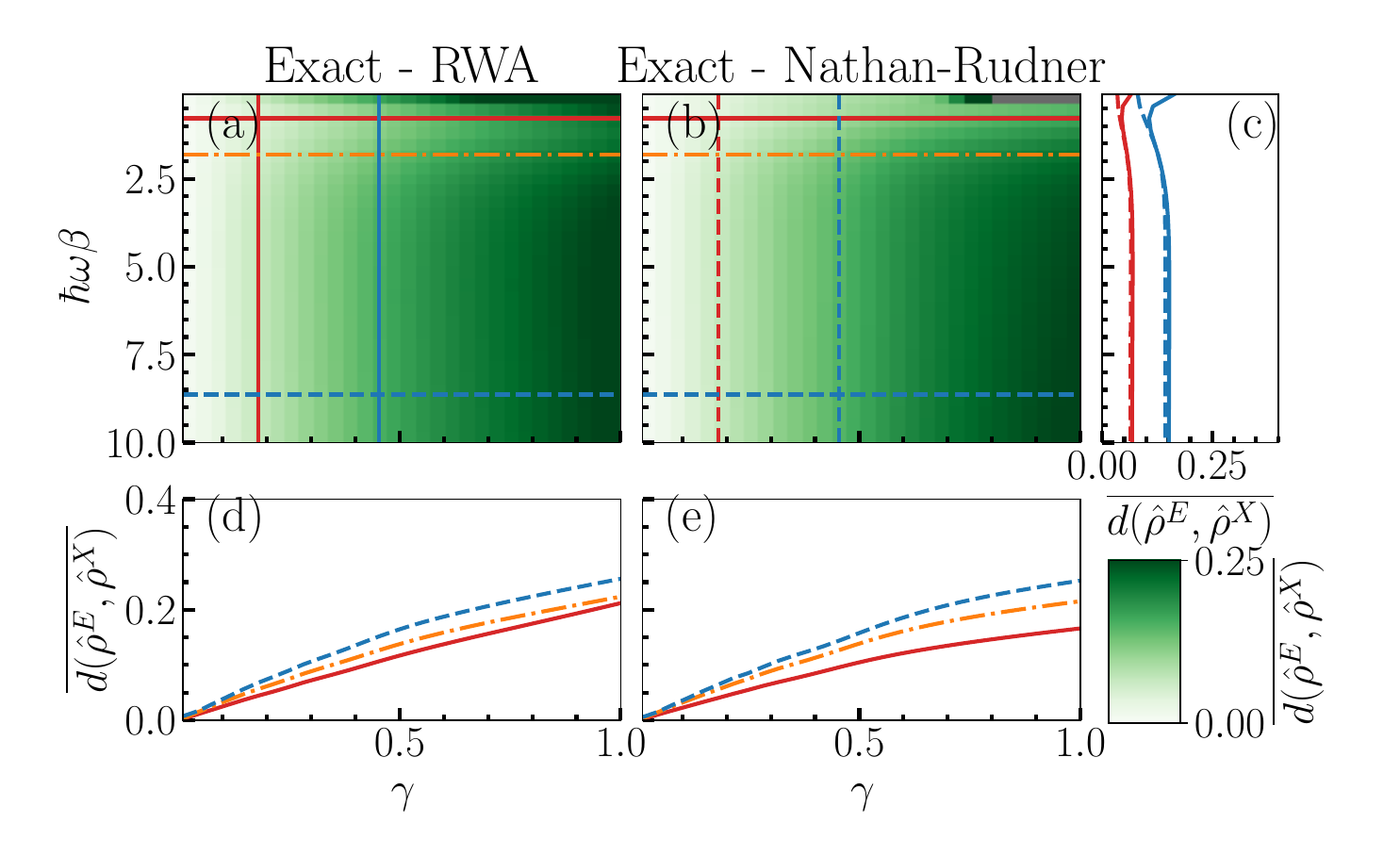}\\%{parameterscan-dynamic-rwa-cutoff5}\\
	\caption[Heatmaps of the average trace distances $d(\hat{\rho}^E,\hat{\rho}^X)$ during time evolution for cutoff $E_c = 5 \hbar \omega$ and varying values of coupling strength $\gamma$ and inverse temperature $\beta$]{Average trace distance in the transient regime  \cref{eq:aver_trace_dist} for cutoff $E_c = 5 \hbar \omega$. (a): Heatmap of $\overline{d(\hat{\rho}^E,\hat{\rho}^{\text{RWA}})}$ for a range of values of coupling strength $\gamma$ and inverse temperature $\beta$. (b): Equivalent Heatmap for $\overline{d(\hat{\rho}^E,\hat{\rho}^N)}$. (c) Plots of $\overline{d(\hat{\rho}^E,\hat{\rho}^X)}$ for specific values of $\gamma$. The corresponding $\gamma$ values for the graphs shown are marked in the heatmaps on the left by vertical lines. (d) and (e): Plots of $\overline{d(\hat{\rho}^E,\hat{\rho}^X)}$ for specific values of $\beta$. The corresponding $\beta$ values for the graphs shown are marked in the heatmaps above by horizontal lines.}
	\label{fig:maxtracedist-timev-rwa}
\end{figure}

Another effect that might partly explain the difference between the dynamical errors of the NRE and the Redfield equation is that the Redfield equation is not of GKSL form. 
While this is a disadvantage, when positivity is actually violated, it might also be an advantage in regimes where this is not the case. 
Namely, the exact Hu-Paz-Zhang equation is equally not of GKSL form (see also Ref.~\cite{beckerOptimalFormTimelocal}), which raises the general question, whether the GKSL form is actually superior to others under all circumstances \cite{tupkaryFundamentalLimitationsLindblad2022,tupkarySearchingLindbladiansObeying2023}.
Therefore, it is natural to compare the exact evolution also to the RWA, which is also of GKSL form.
\cref{fig:maxtracedist-timev-rwa} shows the dynamical error of the RWA and again the dynamical error of the NRE for the same range of values as used in \cref{fig:maxtracedist-timev1}. 
The difference of the dynamical error of the RWA to the error of the NRE is reduced noticeably compared to the difference between the errors of the Redfield and Nathan-Rudner equations. For the dynamics, the RWA and the NRE deliver essentially equal results.

As a first main result of this work we conclude, that for the transient dynamics the Redfield equation is better than the NRE in the parameter regime studied. 
The discrepancy can be mostly attributed to the lack of an explicit time-dependence in the approach by Nathan and Rudner.

\subsection{Steady-state solutions}
\label{sec:steady_state}
In the previous subsection, the transient dynamics of the Nathan-Ruder equation is compared to the transient dynamics of the Redfield equation. 
The focus of this subsection is to analyze steady-state solutions. 
Analogous to the analysis of the transient relaxation dynamics, here, we use the trace distance of the steady-state solutions to the exact equilibrium state of the damped harmonic oscillator as an error measure.
In this section, we focus on the parameter regime in which the Redfield equation still yields a physical steady state. 
This is achieved by choosing the Drude cutoff $E_c=1\, \hbar \omega$ (as compared to $E_c=5\, \hbar \omega$ for the dynamics in \cref{fig:hamiltonian-ex,fig:tracedistance-ex,fig:maxtracedist-timev1,fig:maxtracedist-timev-rwa}).
For the parameters considered in the previous simulations, the Redfield equation violates positivity when both the coupling strength and the inverse temperature take large values (within the parameter range considered).

One of the properties of the Redfield equation is that for ultraweak coupling, when the coupling strength $\gamma$ approaches zero, the steady-state error tends to zero. 
This implies, that in the ultraweak coupling limit, the steady-state solution of the Redfield equation is the canonical Gibbs state of the system.
This can be seen analytically within the perturbation expansion to second order in the coupling strength $\sqrt{\gamma}$ \cite{thingna} and also holds for the NRE \cite{leePerturbativeSteadyStates2022} (see also our discussion in \cref{NR-steady}).
Thus, for both equations, one finds 
 \begin{align}
 \begin{split}
 d(\hat{\rho}^E, \hat{\rho}^X) &= d(\hat{\rho}^E_0 + \gamma \hat{\rho}^E_2, \hat{\rho}^X_0 + \gamma \hat{\rho}^X_2)+\mathcal{O}(\gamma^2) \\ &\xrightarrow{\gamma \rightarrow 0} d(\hat{\rho}^E, \hat{\rho}^X) = d(\hat{\rho}^E_0, \hat{\rho}^X_0) = 0, \label{eq:expansion}
 \end{split}
 \end{align}
with $X = R, \,N$ and $\hat{\rho}^X_0$ and $\hat{\rho}^X_2$ being the zeroth order and second order steady state contributions, respectively.
For finite coupling, the steady-state depends on the details of the system and bath. A detailed analysis for the Redfield equation can be found in Ref.~\cite{thingna}.

\begin{figure}[b!]
	\hspace*{-0.4cm}
	\includegraphics[width=0.55\textwidth]{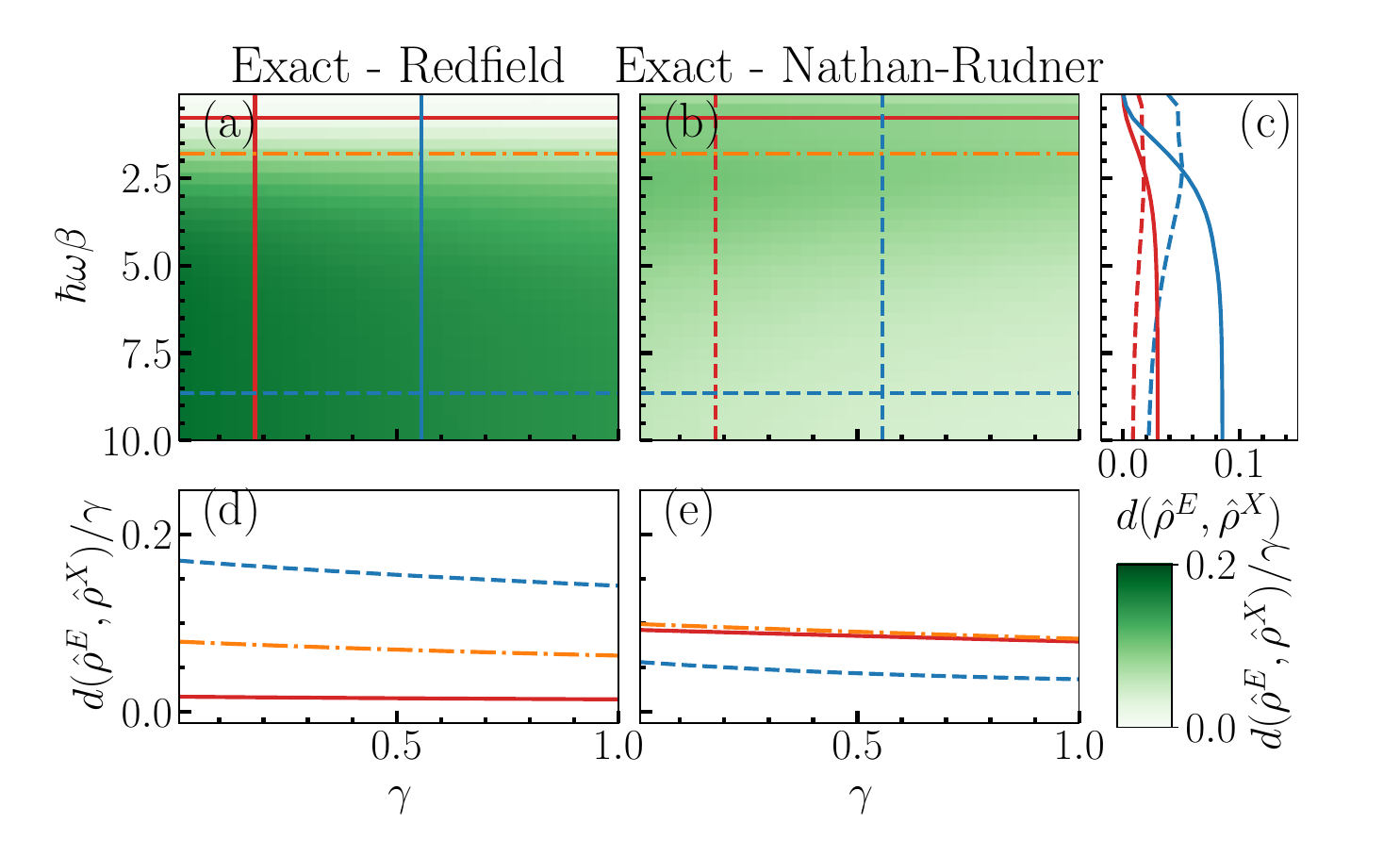}\\  %{parameterscan-stationary-cutoff1}\\
	\caption[Heatmaps of  $d(\hat{\rho}^E,\hat{\rho}^X)/\gamma$ values for steady-state solutions at cutoff $E_c = \hbar \omega$ and varying values of coupling strength $\gamma$ and inverse temperature $\beta$]{$d(\hat{\rho}^E,\hat{\rho}^X)/\gamma$ of steady-state solutions for cutoff $E_c = \hbar \omega$. (a): Heatmap of $d(\hat{\rho}^E,\hat{\rho}^R)/\gamma$ for a range of values of coupling strength $\gamma$ and inverse temperature $\beta$. (b): Equivalent Heatmap for $d(\hat{\rho}^E,\hat{\rho}^N)/\gamma$. (c): Plot of $d(\hat{\rho}^E,\hat{\rho}^X)$ for specific values of $\gamma$. The numerical values have been rescaled with the corresponding value of $\gamma$  compared to the values of the plots on the left. (d) and (e): Plots of $d(\hat{\rho}^E,\hat{\rho}^X)/\gamma$ for specific values of $\beta$. The corresponding $\beta$ values for the graphs shown are marked in the heatmaps above by horizontal lines.}
	\label{fig:paramscan-stat-1st}
\end{figure}

In order to compare the errors beyond the zeroth order, below we plot the scaled errors
\begin{align}
d(\hat{\rho}^E, \hat{\rho}^X)/\gamma  \xrightarrow{\gamma \rightarrow 0} d(\hat{\rho}^E_2, \hat{\rho}^X_2),
\end{align}
which in the limit of $\gamma \to 0$ compares the solutions in second order of the coupling strength. 
For the Redfield equation, the error in second order is expected to be finite \cite{thingna}, which is confirmed in \cref{fig:paramscan-stat-1st}(a) and \cref{fig:paramscan-stat-1st}(d). 
This means that the second-order term of the Redfield steady-state solution does not equal the second-order term of the exact steady-state, i.e.~ $\hat{\rho}^R_2 \neq \hat{\rho}^E_2$.
In \cref{fig:paramscan-stat-1st}(b) and \cref{fig:paramscan-stat-1st}(e), it is shown that the same holds true also for the NRE, i.e.~ $\hat{\rho}^N_2 \neq \hat{\rho}^E_2$. However, we can clearly observe that the Redfield equation performs better for high temperatures, whereas the NRE provides more accurate results for low temperatures.

Let us now focus on the coherences, i.e.~the off-diagonal elements of the density matrix in energy representation. They vanish in zeroth-order perturbation theory, where we obtain the canonical Gibbs state, so that their leading contribution appears in  second order. For the Redfield equation, the second-order coherences are equal to the second-order coherences of the exact equation. 
The reason for the discrepancy in the second order of the steady-state solutions is a discrepancy in the populations (diagonal elements of the density operator) \cite{thingna}. 
To confirm this and to investigate whether such a statement is true also for the NRE, we define the distances of the coherences
\begin{align}
d_{\text{OD}}(\hat{\rho}^E,\hat{\rho}^X) = \sqrt{\sum_{i \neq j}|\rho_{ij}^E -\rho_{ij}^X|^2}.
\end{align}
In order to infer the behaviour in second order, below we again plot the scaled distance
\begin{align}
d_{\text{OD}}(\hat{\rho}^E,\hat{\rho}^X)/\gamma \xrightarrow{\gamma \rightarrow 0} \sqrt{\sum_{i \neq j}|(\rho_{2}^E)_{ij}-(\rho_{2}^X)_{ij}|^2}.
\end{align}
So if $d_{\text{OD}}(\hat{\rho}^E,\hat{\rho}^X)/\gamma$ approaches 0 as $\gamma$ approaches 0, all coherences of $\hat{\rho}^E_2$ and $\hat{\rho}^X_2$ have to be equal, otherwise at least one entry differs.
\begin{figure}[b!]
	\hspace*{-0.4cm}
	\includegraphics[width=0.5\textwidth]{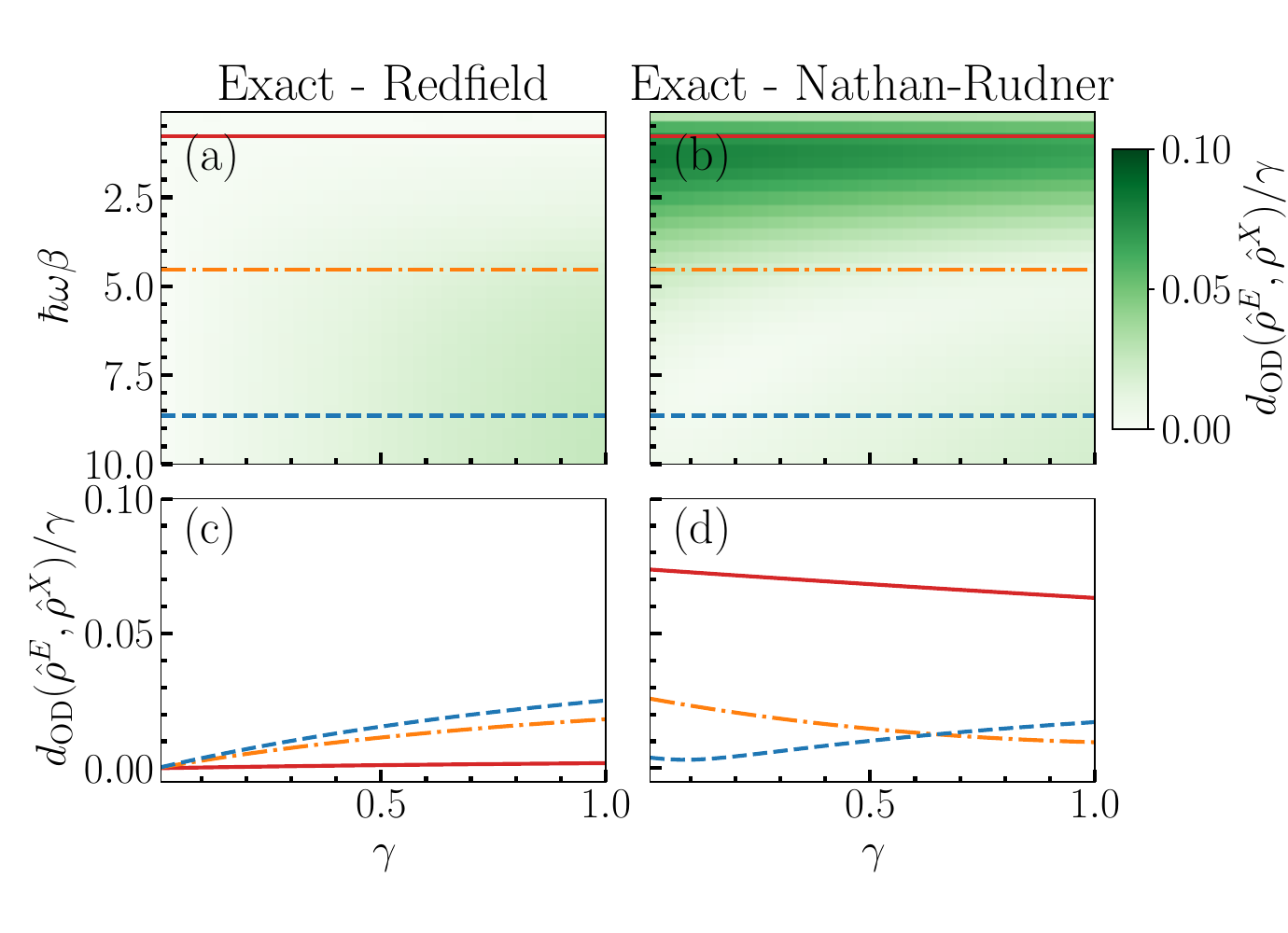}\\   %{parameterscan-stationary-tracedistance-offdiag-1st-cutoff1}\\	
	\caption[Heatmaps of  $d_{\text{OD}}(\hat{\rho}^E,\hat{\rho}^X)/\gamma$ values for steady-state solutions at cutoff $E_c = \hbar \omega$ and varying values of coupling strength $\gamma$ and inverse temperature $\beta$]{$d_{\text{OD}}(\hat{\rho}^E,\hat{\rho}^X)/\gamma$ of steady-state solutions for $E_c = \hbar \omega$. (a): Heatmap of $d_{\text{OD}}(\hat{\rho}^E,\hat{\rho}^R)/\gamma$ for a range of values of $\gamma$ and $\beta$. (b): Equivalent Heatmap for $d_{\text{OD}}(\hat{\rho}^E,\hat{\rho}^N)/\gamma$. (c) and (d): Plots of $d_{\text{OD}}(\hat{\rho}^E,\hat{\rho}^X)/\gamma$ for specific values of $\beta$. The corresponding $\beta$ values for the graphs shown are marked in the heatmaps above by horizontal lines.}
	\label{fig:paramscan-stat-offdiag}
\end{figure}
\noindent As shown in \cref{fig:paramscan-stat-offdiag}(a) and \cref{fig:paramscan-stat-offdiag}(c), for the Redfield solution the quantity $d_{\text{OD}}(\hat{\rho}^E,\hat{\rho}^R)/\gamma$ approaches zero in the limit of $\gamma\to 0$.
This confirms the prediction that the coherences of the Redfield steady state agree with the exact solution up to second order in the coupling.
Furthermore, this is the case for all temperatures. 
In contrast, in \cref{fig:paramscan-stat-offdiag}(b) and \cref{fig:paramscan-stat-offdiag}(d), for the NRE the quantity $d_{\text{OD}}(\hat{\rho}^E,\hat{\rho}^N)/\gamma$ approaches finite values for all considered temperatures.
Consequently, in second order of the coupling, the co\-he\-ren\-ces of the Nathan-Rudner steady-state solution deviate from the exact equilibrium state.
It was recently discussed in Ref.~\cite{tupkaryFundamentalLimitationsLindblad2022} that such a deviation can give rise to a spurious violation of local conservation laws (like energy or particle number conservation in the bulk of a system coupled to an environment at its edges).
To address the problem of an inconsistent steady state of the NRE in second order of the coupling, which was also pointed out in Ref.~\cite{NR-comment,nathanResponseCommentUniversal2020}, Nathan and Rudner proposed an additional transformation that is applied after integrating the NRE and that is supposed to resolve these issues \cite{nathanQuantifyingAccuracySteady2024}. 
However, the modification comes with an considerable effort and is omitted in our consideration here.

In \cref{fig:paramscan-stat-offdiag}(c), we can see that the error in the coherences of the NRE becomes particularly pronounced in the limit of high temperatures. And also in \cref{fig:paramscan-stat-1st}(c), we can see clear deviations from the exact solution.
In the limit of ultrahigh temperatures, $\hbar \omega \beta \ll 1$, for states with energy well below the temperature, $\epsilon_n \ll 1/ \hbar \omega \beta$, the density matrix approaches the maximally mixed state with equal populations and vanishing coherences.
In this regime, the bath correlation is short lived and the Redfield equation as well as the RWA and other Lindbladian approximations, e.g., in Ref.~\cite{tobias}, are valid descriptions. 
However, for the steady state of the NRE, in \cref{fig:paramscan-stat-1st}(c), in the high-temperature limit the errors appear to be finite.
This discrepancy is likely be related to the fact that the NRE does not accurately capture the off-diagonal elements.

\begin{figure}[t!]
	\hspace*{-0.4cm}
	\includegraphics[width=0.55\textwidth]{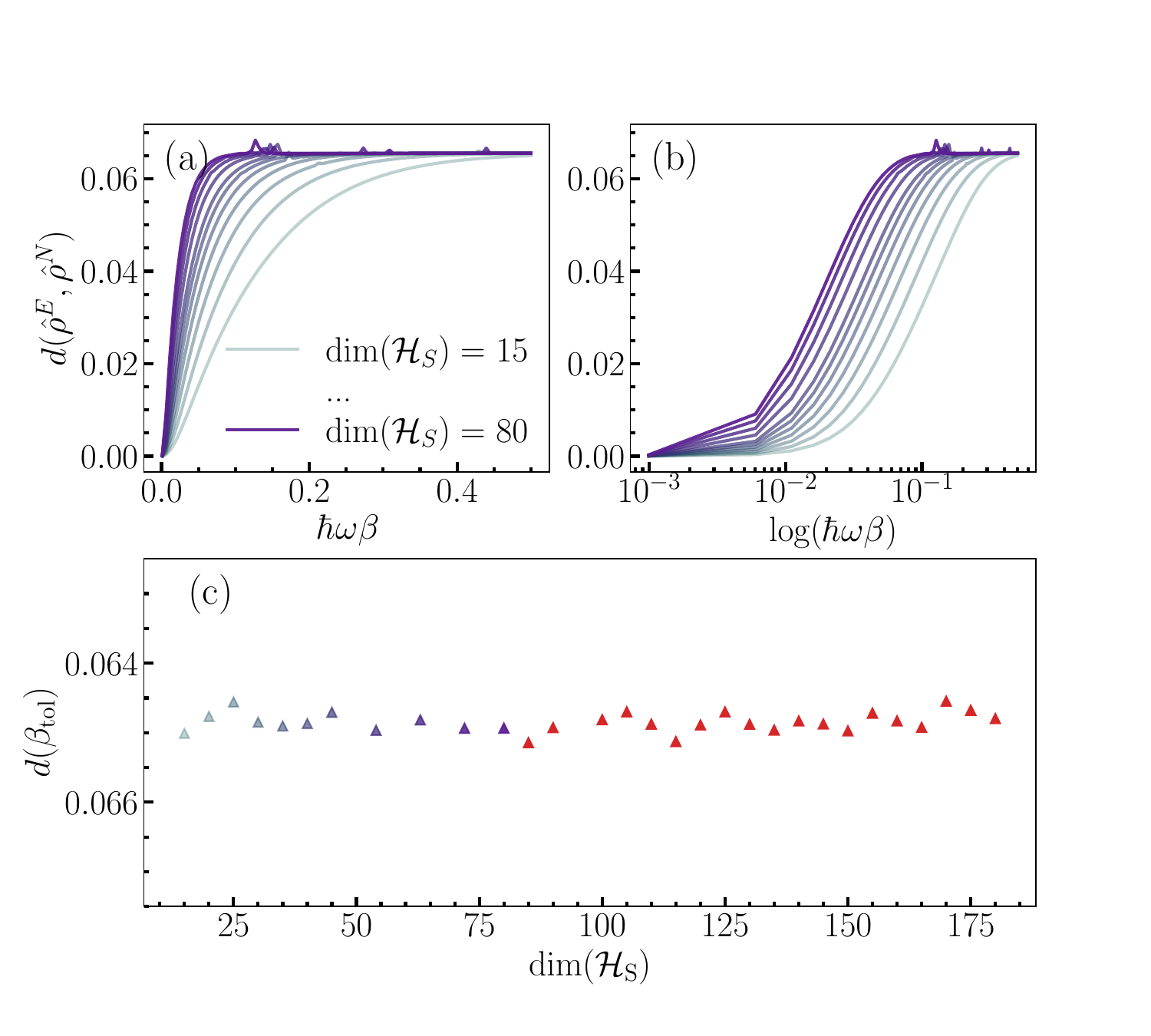}\\ %{stationary-tracedistance-gamma0.8-cutoff1}\\	
	\caption[Steady-state error $d(\hat{\rho}^{E},\hat{\rho}^{E})$ of the NRE for increasing Hilbert space dimension of the system for $\gamma = 0.8$ and varying values of $\beta$]{(a): Steady-state error $d(\hat{\rho}^{E},\hat{\rho}^{E})$ of the NRE for increasing Hilbert space dimension $\text{dim}\big( \mathcal{H}_{\mathcal{S}}\big)$ and varying values of $\beta$, where $\gamma = 0.8$. $\beta_{\text{tol}}$ is the value where the tolerance of the numerical calculation is reached. (b): same plot as (a) with logarithmic inverse temperature scale. (c): steady-state error $d(\beta_{\text{tol}})$ of NRE at tolerance temperature $\beta_{\text{tol}}$ as a function of Hilbert space dimension $\mathrm{dim}(\mathcal{H}_\mathrm{S})$.}
	\label{fig:truncation}
\end{figure}

In order to investigate the behaviour in the high-temperature regime in more detail, in \cref{fig:truncation}(a), we plot the steady-state error of the NRE as a function of the inverse temperature and for increasing truncations of the Hilbert space dimension. While this more detailed plot of the ultrahigh-temperature regime seems to imply that the steady-state error of the NRE goes to zero, the rapid drop off of the steady-state error for higher dimensions of the Hilbert space turns out to be an artifact of the truncation of the Hilbert space. 
For the steady state, the energy at which the state space is truncated should be well above both the temperature and the system-bath coupling. 
To ensure the accuracy of our numerical simulations, we estimate a tolerance value for the sum of the populations of the highest ten percent of energy states, which should not be exceeded. 
This also constraints the maximal considered temperature, or equivalently minimal inverse temperature. 
We set the tolerance to be 0.1\% and call the inverse temperature, where the tolerance is reached $\beta_{\text{tol}}$. For smaller inverse temperatures $\beta < \beta_\mathrm{tol}$ the highest energy states get populated beyond our tolerance.
In \cref{fig:truncation}(c), we plot the corresponding steady-state error $d(\beta_{\text{tol}})$ as a function of the truncated Hilbert space dimension $\mathrm{dim}(\mathcal{H}_\mathrm{S})$, where we use the same color for the triangles as for the corresponding curves in \cref{fig:truncation}(a). For larger Hilbert space dimension, we use red triangles.
\cref{fig:truncation}(c) indicates that for increasing Hilbert space dimension, and therefore decreasing $\beta_\mathrm{tol}$, the steady-state error tends to a finite value.
This implies, that the NRE does not reproduce the exact steady state in the ultrahigh temperature limit.

\section{Conclusions}
In this work, we study the regime of validity of the Nathan-Rudner quantum master equation \cite{nathan-rudner} both for the transient dynamics and its steady state. 
By benchmarking it against the exact solution of the damped harmonic oscillator in a thermal bath, we show that for the transient dynamics the Redfield equation outperforms the NRE, whereas the NRE delivers similar results to the quantum optical master equation.
For the steady state and for low temperatures, the NRE performs better than the Redfield equation, which in this regime can fail due to strong violation of positivity.
In turn, for high temperatures the Redfield equation becomes exact, which appears not to be the case for the NRE.
However, it has been shown \cite{leePerturbativeSteadyStates2022} that like the Redfield equation also the NRE provides the correct canonical Gibbs state in the limit of ultraweak coupling. 
In conclusion, for the steady state the NRE can provide a very useful tool for the investigation of open quantum systems, especially at low temperatures, where for the paradigmatic damped oscillator model it performs better than the Redfield equation. 
Whether the latter is true in general should be the focus of future studies.

One interesting question for future research is, whether it is possible to find GKSL master equations similar to the NRE that are time dependent, so that at low temperature and beyond ultraweak coupling they describe not only the steady state, but also the transient dynamics.
Another one is, whether it is possible to construct such a GKSL master equation, which correctly describes the steady state in second order system bath coupling, as the recently proposed non-GKSL canonically consistent master equation \cite{beckerCanonicallyConsistentQuantum2022}.

\begin{acknowledgments}
We thank, Frederik Nathan, Archak Purkayastha, and Mark Rudner for their helpful comments.
This research was funded by the Deutsche Forschungsgemeinschaft (DFG) via the Research Unit FOR 2414 under the Project No. 277974659. 
\end{acknowledgments}

\bibliography{ref}

\appendix
\section{Transition to Schrödinger picture}
\label{schroedinger}
Here we will show, that setting $t_0 = -\infty$ in \cref{eq:start} and setting $t_0 = 0$ while changing the upper integration boundary to $\infty$ leads to the same Redfield equation in the Schrödinger picture. Expressing the left-hand-side of \cref{eq:start} with respect to the density operator in the Schrödinger picture, one finds
\begin{align}
\begin{split}
&\frac{d}{dt}(\exp(i\hat{H}_St/\hbar)\hat{\rho}(t)\exp(-i\hat{H}_St/\hbar))  \\&= \exp(i\hat{H}_St/\hbar)\left(\frac{d}{dt}\hat{\rho}(t) + \frac{i}{\hbar}[\hat{H}_S, \hat{\rho}(t)]\right)\exp(-i\hat{H}_St/\hbar). 
\end{split}
\end{align}
For the integrand on the right-hand side, one finds 
\begin{align}
[\hat{S}_{t}, \hat{S}_{\tau} \hat{\rho}_I(t)] = \exp(i\hat{H}_St/\hbar)[\hat{S}, \hat{S}_{\tau-t} \hat{\rho}(t)]\exp(-i\hat{H}_St/\hbar).
\end{align}
This leads to
\begin{align}
\frac{d}{dt}\hat{\rho}(t) = -\frac{i}{\hbar}[\hat{H}_S, \hat{\rho}(t)] + \int_{t_0}^{t}d\tau C(t-\tau)[\hat{S}, \hat{S}_{\tau - t} \hat{\rho}(t)]+\text{H.c.}.
\end{align}
Making the variable transformation from $\tau$ to $t-\tau$ results in
\begin{align}
\frac{d}{dt}\hat{\rho}(t) = -\frac{i}{\hbar}[\hat{H}_S, \hat{\rho}(t)] + \int_{0}^{t-t_0}d\tau C(\tau)[\hat{S}, \hat{S}_{-\tau} \hat{\rho}(t)]+\text{H.c.}.
\end{align}
The upper integration boundary is only dependent on the difference $t-t_0$. For $t_0 = -\infty$, this difference is $\infty$. The same is the case if we set $t$ in the upper integration boundary to $\infty$ and $t_0 = 0$. The lower integration boundary is always 0.

\section{Steady-state solution of Nathan-Rudner Equation in the ultraweak coupling limit}
\label{NR-steady}
In the main text, in \cref{benchmark}, we have analyzed the validity of the NRE numerically for the damped harmonic oscillator coupled to an Ohmic bath with Drude cutoff. 
The numerical analysis indicates, that for this specific model the NRE yields the exact steady state in the ultraweak coupling limit, which is equal to the canonical Gibbs state of the system. 
In this section, we show analytically that this is true for arbitrary systems, just as it is the case for the Redfield equation.
With this, we repeat the analysis of Refs.~\cite{NR-comment,leePerturbativeSteadyStates2022} where the authors obtained the steady-state solution of the NRE up to second order in the coupling strength.
To do that, we start from the steady-state solution of the Nathan-Rudner \cref{eq:NR}
\begin{align}
	0 = -\frac{i}{\hbar}[\hat{H}_S + \gamma  \hat{\tilde{\Lambda}}^{N}, \hat{\rho}^{N}] -\gamma\bigg(\frac{1}{2}\{\hat{\tilde{L}}^{\dagger}\hat{\tilde{L}}, \hat{\rho}^{N}\} + \hat{\tilde{L}}\hat{\rho}^{N}\hat{\tilde{L}}^{\dagger}\bigg), \label{eq:NR-stat}
\end{align}
where we have indicated explicitly that the generator is of second order in the coupling $\sqrt{\gamma}$ by writing, $\hat{L} = \sqrt{\gamma}\hat{\tilde{L}}$ and $\hat{\Lambda}^{N} = \gamma\hat{\tilde{\Lambda}}^{N}$.
For the steady state we perform a perturbative expansion up to second order in the coupling, $\hat{\rho}^{N} \simeq \hat{\rho}^{N}_{0} + \gamma \hat{\rho}^{N}_{2}$, and arrange the terms in orders of $\gamma$.
We arrive at
\begin{align}
	\begin{split}
		0 =& -\frac{i}{\hbar}[\hat{H}_S, \hat{\rho}^{N}_{0}] + \gamma  \bigg(-\frac{i}{\hbar}\big([\hat{H}_S, \hat{\rho}^{N}_{2}] +[\hat{\tilde{\Lambda}}^{N}, \hat{\rho}^{N}_{0}]\big) \\ &-\frac{1}{2}\{\hat{\tilde{L}}^{\dagger}\hat{\tilde{L}}, \hat{\rho}^{N}_{0}\} + \hat{\tilde{L}}\hat{\rho}^{N}_{0}\hat{\tilde{L}}^{\dagger}\bigg) + \mathcal{O}(\gamma^2) , \label{eq:NR-gamma}
	\end{split}
\end{align}
Since $\gamma$ can take any (small) value, the zeroth and second order have to vanish independently in \cref{eq:NR-gamma}. 
The zeroth-order contribution reads
\begin{align}
	\begin{split}
		0 &= [\hat{H}_S, \hat{\rho}^{N}_{0}]  \\ &= \big[\sum_{n}\epsilon_{n}\ketbra{n}{n} \sum_{lk} \langle l| \hat{\rho}^{N}_{0}|k\rangle \ketbra{l}{k}\big] \\
		&= \sum_{lk}\langle l| \hat{\rho}^{N}_{0}|k\rangle \, (\epsilon_{l}-\epsilon_{k})\ketbra{l}{k},
	\end{split}
\end{align}
where $\ket{k}$ and $\epsilon_k$ denote an energy eigenstate and its energy, respectively. It follows that $\langle l| \hat{\rho}^{N}_{0}|k\rangle \, (\epsilon_{l}-\epsilon_{k})=0$ for all pairs $l$ and $k$. 
For $l = k$ this is always the case because $\Delta_{ll} = \epsilon_{l}-\epsilon_{l} = 0$. 
This line of reasoning can also be used to show that the diagonal entries of the commutator $\langle l |[\hat{H}_S, \hat{\rho}^{N}_{2}]|l \rangle =0$ are zero.
For $l\ne k$, when $\Delta_{lk}\ne 0$, it follows that $\langle l| \hat{\rho}^{N}_{0}|k\rangle = 0$.  
Therefore, $\hat{\rho}^{N}_{0}$ is diagonal in the energy eigenbasis and can be written as $\hat{\rho}^{N}_{0} = \sum_{n}p_n\ketbra{n}{n}$, with populations $p_n$. 
In second order, i.e. in linear order with respect to $\gamma$, using the Lamb shift in \cref{eq:NR-def}, the commutator can be written as 
\begin{align}
	[\hat{\tilde{\Lambda}}^{N}, \hat{\rho}^{N}_{0}] =  \sum_{mlk}\tilde{f}(\Delta_{ml},\Delta_{km})S_{lm}S_{mk}(p_k-p_l)\ketbra{l}{k},
\end{align}
where $ \gamma\tilde{f}(\Delta_{ml},\Delta_{km}) = f(\Delta_{ml},\Delta_{km})$. 
Again, the diagonal entries of $[\hat{\tilde{\Lambda}}^{N}, \hat{\rho}^{N}_{0}]$ are zero because $p_k-p_l$ vanishes for $l = k$. The remaining parts of the second-order term can be written as
\begin{align}
	\{\hat{\tilde{L}}^{\dagger}\hat{\tilde{L}}, \hat{\rho}^{N}_{0}\} = 4\pi^2\sum_{ilk}\tilde{h}(\Delta_{li})\tilde{h}(\Delta_{lk})S_{il}S_{lk}(p_k+p_i)\ketbra{i}{k} \label{eq:L1}
\end{align}
and
\begin{align}
	\hat{\tilde{L}}\hat{\rho}^{N}_{0}\hat{\tilde{L}}^{\dagger} = 4\pi^2\sum_{lkj}p_k\tilde{h}(\Delta_{lk})
	\tilde{h}(\Delta_{jk})S_{jk}S_{lk}\ketbra{l}{j}, \label{eq:L2}
\end{align}
where the system representation from \cref{eq:NR-def} and $\sqrt{\gamma}\tilde{h}(\Delta_{lk}) = h(\Delta_{lk})$ are used. Now we focus on the diagonal elements of the second-order contribution, which yield an algebraic equation for the populations $p_i$.
Combining \cref{eq:L1,eq:L2} and using the fact that the diagonal elements of the commutators in the second-order term of \cref{eq:NR-gamma} are zero, one obtains
\begin{align}
	\begin{split}
		&\langle i |\big(-\frac{1}{2}\{\hat{\tilde{L}}^{\dagger}\hat{\tilde{L}}, \hat{\rho}^{N}_{0}\} + \hat{\tilde{L}}\hat{\rho}^{N}_{0}\hat{\tilde{L}}^{\dagger}\big)| i\rangle \\ &=  2\pi^2\sum_{l}\big(2p_lh(\Delta_{il})^2-2p_ih(\Delta_{li})^2\big). \label{eq:pauli}
	\end{split}
\end{align}
Equation (\ref{eq:pauli}) has the form of a rate equation, where $h(\Delta_{il})^2$ are the transition rates.
We can use $h(\Delta)=\sqrt{\frac{1}{2\pi}\frac{J(\Delta)/\hbar}{e^{\beta \Delta}-1}}$, which results in the detailed-balanced condition \cite{breuer}
\begin{align}
	\frac{h(\Delta_{li})^2}{h(\Delta_{il})^2} = -\frac{e^{\beta \Delta_{il}}-1}{e^{\beta \Delta_{li}}-1} = e^{-\beta \Delta_{li}},\label{eq:balance}
\end{align}
where $J(-\Delta) = -J(\Delta)$ was employed. Inserting \cref{eq:balance}, \cref{eq:pauli} can be simplified further. Since the resulting expression has to be equal to 0, one obtains
\begin{align}
	\sum_{l}h(\Delta_{il})^2\big(p_l-p_ie^{-\beta \Delta_{li}}\big) = 0.
\end{align}
As the rates are non-negative and the equation has to be equal to 0 for all $p_i$, the condition
\begin{align}
	\frac{p_l}{p_i} = e^{-\beta \Delta_{li}}
\end{align}
has to be satisfied. This can be rewritten as $p_l = p_ie^{\beta(\epsilon_i-\epsilon_l)}$, with $\epsilon_i$ and $\epsilon_l$ being eigenenergies of the system. With the normalization condition $\sum_{l} p_l =1$ it follows, that $\hat{\rho}^{N}_{0} = \exp(-\beta \hat{H}_S)/Z_S$, where $Z_S$ is the partition function of the system. 

At finite coupling, the steady state of the NRE does not correspond to the canonical Gibbs state anymore \cite{NR-comment,nathanResponseCommentUniversal2020}. However, in this regime also the thermal state of system and bath, does not correspond to the canonical Gibbs state of the system, but to a state that can be expressed as a Gibbs state of the so-called Hamiltonian of mean force. As we have discussed in the main text, in \cref{sec:steady_state}, neither the Redfield nor the Nathan-Rudner steady state correspond to this exact steady state at finite coupling (though the Redfield equation at least possesses the correct second-order coherences \cite{thingna}).

\section{Time-independent generator}
\label{generator}
In this section, we explicitly state the time-independent generators for the Redfield equation and RWA 
for the damped harmonic oscillator in an Ohmic bath with Drude cutoff, where $\hat{\mathbb{S}}_{t}$ from \cref{eq:Redfield} is replaced with $\hat{\mathbb{S}}_{\infty}$, which is the limit of $\hat{\mathbb{S}}_{t}$ for $t$ approaching infinity.

At a microscopic level, the bath is modeled as a continuum of harmonic oscillators. Therefore, the bath Hamiltonian reads
$\hat{H}_B = \sum_{n}\hbar\omega_n(\hat{b}_n^{\dagger}\hat{b}_n + \frac{1}{2})$, where $\hat{b}_n^{\dagger}$ and $\hat{b}_n$ are creation and annihilation operators of the harmonic oscillators in the bath.

The one-dimensional position operator $\hat{x}$ of the system will be used as the coupling operator $\hat{S}$. The other coupling operator $\hat{B}$ will be modeled as $\hat{B} = \sum_{n}\kappa_n \hat{x}_n$, with $\hat{x}_n$ being the position operators of the harmonic oscillators of the bath and $\kappa_n$ the corresponding coupling strengths \cite{breuer}.
With these definitions and by assuming that the bath is in the canonical Gibbs state $\hat{\rho}_B = \exp(-\beta \hat{H}_B)/\Tr_B(\exp(-\beta \hat{H}_B))$ at inverse temperature $\beta$, the bath correlation function $C(\tau)$ can be expressed with the spectral density $J(\Omega)=\sum_n|\kappa_n|^2\delta(\Omega-\hbar\omega_n)$ \cite{tobias, breuer}, so that 
\begin{align}
	C(\tau) = \int_{-\infty}^{\infty}\frac{d\Omega}{\hbar} \frac{J(\Omega)/\hbar}{e^{\beta \Omega}-1}e^{i\Omega \tau/ \hbar}.
\end{align}
To calculate the matrix elements of the convoluted operator $\hat{\mathbb{S}}_{\infty}$, it is convenient to introduce the eigenbasis of the system. The matrix elements are then given by $(\hat{\mathbb{S}}_{\infty})_{lk} = G_{\infty}(\Delta_{lk})x_{lk}$
with the energy difference $\Delta_{lk}$ between the eigenstates $\ket{l}$ and $\ket{k}$ and the position operator matrix elements $x_{lk}$. $G_{\infty}(\Delta)$ is the half-sided Fourier transform of the bath correlation function $C(\tau)$. Rewriting and changing the order of integration results in
\begin{align}
	G_{\infty}(\Delta) = \frac{1}{\hbar^2} \int_{-\infty}^{\infty}\frac{d\Omega}{\hbar} \frac{J(\Omega)/\hbar}{e^{\beta \Omega}-1}\int_{0}^{\infty}d\tau e^{i(\Omega -\Delta)\tau/\hbar}.
\end{align}
By making use of the Sokhotski–Plemelj formula  \cite{merzbacher}
\begin{align}
	\int_{0}^{\infty}d\tau e^{i(\Omega -\Delta)\tau/\hbar} = \hbar \left( \pi \delta (\Omega-\Delta) -i \mathcal{P}\frac{1}{\Omega -\Delta}\right),
\end{align}
where $\delta(.)$ is the Dirac-delta function and $\mathcal{P}$ the principal value, $G_{\infty}(\Delta)$ can be divided into real and imaginary part,
\begin{align}
	\begin{split}
		G_{\infty}(\Delta)
		&= \pi \frac{ J(\Delta)/\hbar}{e^{\beta \Delta}-1} -i \mathcal{P}\int_{-\infty}^{\infty}d\Omega\frac{1}{\Omega -\Delta}\frac{ J(\Omega)/\hbar}{e^{\beta  \Omega}-1} \\ &= G^r_{\infty}(\Delta) + iG^i_{\infty}(\Delta).
	\end{split}
\end{align}
The spectral density $J(\Delta)$ is now assumed to be a smooth function. For our analysis, we use an Ohmic bath with a Drude cutoff function (\cref{eq:drude}).
Using this model, an analytic expression for $G_{\infty}^i(\Delta)$ can be obtained \cite{tobias}
\begin{figure}[t]
	%\hspace*{-1cm}
	\includegraphics[width=0.35\textwidth]{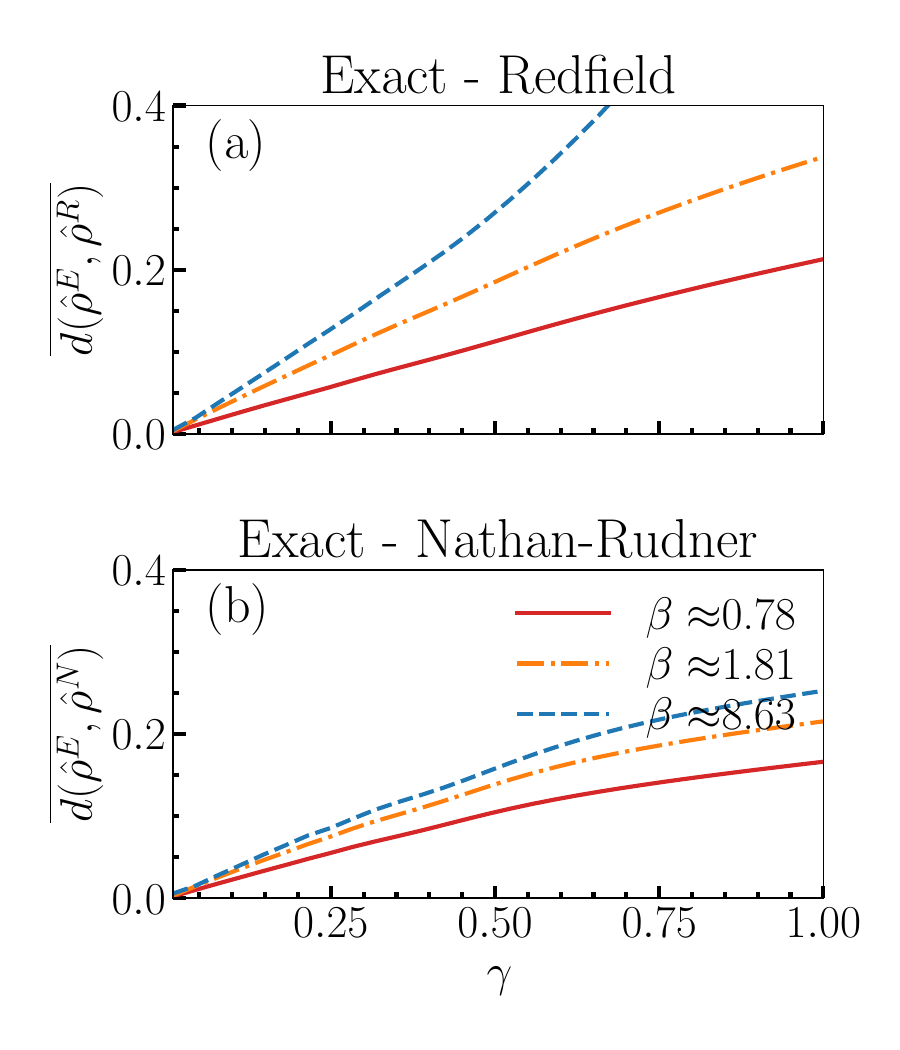}\\    %{parameterscan-dynamic-redfield-tind-cutoff5}\\	
	\caption[Dynamical error $d(\hat{\rho}^E,\hat{\rho}^X)$ of the different master equations for $E_c = 5 \hbar \omega$ for a range of values of $\gamma$ using examplary values for $\beta$.]{Dynamical error $d(\hat{\rho}^E,\hat{\rho}^X)$ of the different master equations for $E_c = 5 \hbar \omega$ for a range of values of $\gamma$ using examplary values for $\beta$. (a): Plot of $d(\hat{\rho}^E,\hat{\rho}^R)$ for specific values of $\beta$. (b): Plot of $d(\hat{\rho}^E,\hat{\rho}^N)$ for specific values of $\beta$.}
	\label{fig:timeindependent}
\end{figure}
\begin{align}
	\begin{split}
		G_{\infty}^i(\Delta) &= \frac{-\gamma \Delta^2 E_c}{2\hbar(E_c^2 + \Delta^2)} \\ &+ (\Delta/\hbar)\gamma\bigg(\frac{-E_c^2}{2(E_c^2 + \Delta^2)}\cot(\beta E_c/2) \\ &+\frac{2}{\beta}\sum_{l=1}^{\infty}\frac{\nu_l}{(\Delta^2+\nu_l^2)(1-\nu_l^2/E_c^2)}\bigg),
	\end{split}
\end{align}
with the Matsubara energies defined as $\nu_l=2\pi l/\beta$.
The spectral density $J(\Delta)$ is now assumed to be a smooth function. 

To see how the accuracy of the Redfield equation changes when we use the time-independent generator, we have reproduced \cref{fig:maxtracedist-timev1}(d) in \cref{fig:timeindependent}(a), this time using the time-independent generator. The figure shows the dynamical error of the Redfield equation for varying values of $\gamma$, using exemplary values of $\beta$. Again, we have included the same plot for the Nathan-Rudner equation using the same exemplary values for $\beta$, which can be seen in \cref{fig:timeindependent}(b).
Compared to the time-dependent case, the Nathan-Rudner equation delivers comparable results to the Redfield equation and even outperforms it in the low-temperature regime. This implies, that much of the advantage that the Redfield equation has in reproducing the transient dynamics of the system comes from the fact that it is not limited to the time-dependent case.
\end{document}